\documentstyle[preprint,aps]{revtex}
\begin{document}
\draft
\title{
THE SINGLE-PARTICLE SPECTRAL FUNCTION OF $^{16}{\rm O}$     }
\author{H. M\"uther}
\address{
Institut f\"ur Theoretische Physik, Universit\"at T\"ubingen,\\
Auf der Morgenstelle 14, 72076 T\"ubingen, Germany  }
\author{W.H. Dickhoff}
\address{
Department of Physics, Washington University,\\
St. Louis, MO 63130, USA}
\date{\today}
\maketitle
\begin{abstract}
The influence of short-range correlations on the $p$-wave single-particle
spectral function in $^{16}{\rm O}$ is studied as a function of energy.
This influence, which is represented by the
admixture of high-momentum components, is found to be
small in the $p$-shell quasihole wave functions.
It is therefore unlikely that studies of quasihole momentum
distributions using the $(e,e'p)$ reaction will reveal a significant
contribution of high momentum components.
Instead, high-momentum
components become increasingly more dominant at higher excitation
energy.
The above observations are consistent with the energy distribution
of high-momentum components in nuclear matter.
\end{abstract}


The influence of short-range correlations in finite nuclei has been
studied theoretically by calculating the momentum distribution
in the ground state of a particular nucleus\cite{zabol,orden,benha,ji,string}.
These results clearly show that for momenta above 400 MeV/c short-range
and tensor correlations completely dominate the momentum distribution.
The momentum distribution for a given $lj$-combination is given by
\begin{equation}
n_{lj}(k) = \left\langle \Psi_0 \right\vert a^{\dag}_{klj} a_{klj}
\left\vert \Psi_0 \right\rangle .
\label{eq:nljk}
\end{equation}
The total momentum distribution is obtained by summing over all
$lj$ combinations multiplying each contribution with the relevant
degeneracy factor.
It is possible to rewrite Eq.~(\ref{eq:nljk}) by inserting a complete
set of $A-1$ particle states with the result
\begin{equation}
n_{lj}(k) = \sum_n \bigl| \left\langle \Psi_n \right| a_{klj}
\left| \Psi_0 \right\rangle \bigr|^2 .
\label{eq:nljke}
\end{equation}

In a simple mean-field description this sum is exhausted by the transition
from the ground state of the $A$ particle system to the ground state
of the $A-1$ particle system when the $lj$ combination corresponds
to the last occupied single-particle level.
In that case Eq.~(\ref{eq:nljke}) represents
nothing but the square of the corresponding single-particle wave
function in momentum space.
When correlations beyond the mean field are present, this is no longer
true, although the ground state to ground state transition might still
dominate at least for small momenta.
This has been observed in Ref.~\cite{tadok} for the momentum distribution
of $^3{\rm He}$.
In this work the ground state to ground state transition was calculated
for both $^3{\rm He}$ and $^4{\rm He}$.
For $^4{\rm He}$ some sensitivity to short-range correlations was observed
in the second maximum.
{}From a comparison to the total momentum distribution of
$^3{\rm He}$, which is calculated in Ref.~\cite{ciofi},
it becomes clear, however, that the ground state to ground state
contribution to Eq.~(\ref{eq:nljke}) contributes only an insignificant
fraction of the high momentum components.
{}From Eq.~(\ref{eq:nljke})
one may thus infer that the high-momentum components must come from the
contribution of excited states in the $A-1$ system.
In other words, one requires knowledge of the complete energy dependence
of the nucleon hole spectral function
\begin{equation}
S_{lj}(k,E) = \sum_n  \bigl| \left\langle \Psi_n \right| a_{klj}
\left| \Psi_0 \right\rangle \bigr|^2 \delta (E-(E^A_0-E^{A-1}_n)) .
\label{eq:sljke}
\end{equation}
By integrating $S_{lj}(k,E)$ from $- \infty$ to $\epsilon^-_F=E^A_0
-E^{A-1}_0$, which represents the energy difference between the
corresponding ground states, one obtains the contribution from
this particular $lj$ combination to the total momentum distribution.
Clearly $S_{lj}(k,E)$ contains the information on the location of
high-momentum components which can be studied in the $(e,e'p)$
reaction.

A recent proposal to study short-range correlations with the $(e,e'p)$
reaction focuses on the low-lying discrete transitions
at high missing momentum\cite{lapik}.
This proposal has been inspired by the work of Ref.~\cite{lewar} for
$^3{\rm He}$ droplets of a finite number of atoms.
In this work the corresponding coordinate space contribution
to Eq.~(\ref{eq:nljke}) from the ground state to ground state transition
was evaluated.
A simple procedure was developed to obtain the amplitude
$\left\langle \Psi_n \right| a_{rlj} \left| \Psi_0 \right\rangle$,
usually referred to as the quasihole wave function,
from a corresponding mean-field wave function.
In Ref.~\cite{mawam} a phenomenological prescription was developed
to study the change from standard Woods-Saxon wave functions to the
corresponding quasihole wave function for nuclei.
A general discussion of quasihole (quasiparticle) properties and a
many-body analysis based on experimental information is available
in Ref.~\cite{mahau}.
Based on the work in Refs.~\cite{mawam,mahau} one obtains a suppression
of the mean-field wave function in the nuclear
interior which results in a corresponding
quasihole wave function with high-momentum components, which are sensitive
to this suppression.
Whether these high-momentum components follow
from the inclusion of short-range correlations induced by a realistic
nucleon-nucleon interaction is not clear however.

In order to study this question and to elucidate the presence
of high-momentum components in nuclei, the nucleon hole spectral
function for the $p$-states in $^{16}{\rm O}$ has been calculated
in a complete energy domain.
Short-range and tensor correlations have been evaluated explicitly for the
finite system under study.
Although the nucleon hole spectral function has been carefully studied
in nuclear matter starting with the work of Refs.~\cite{ramos,benff}
(see also a review in Ref.~\cite{dickh}), no complete
microscopic calculations are available for nuclei heavier than $^3{\rm He}$
\cite{diepe}.
The relevant method for the present study has been developed in
Ref.~\cite{borro}.
In Ref.~\cite{borro} the nucleon self-energy was calculated in $^{16}{\rm O}$
using a $G$-matrix interaction calculated from a realistic nucleon-nucleon
interaction.
A Hartree-Fock like contribution was identified which is obtained by using
a real $G$-matrix, calculated in nuclear matter at an
appropriate starting energy and density, in the corresponding
Hartree-Fock diagram for the self-energy in $^{16}{\rm O}$.

The imaginary part of the self-energy is obtained by calculating the
relevant second order diagrams in this $G$-matrix interaction which
contain the two-particle-one-hole and two-hole-one-particle terms
appropriate for this nucleus.
The former term is responsible for the depletion of strength, which in mean
field is located below the Fermi energy, to high energy.
The latter term is essential for the accumulation of single-particle
strength below the Fermi energy from states (in particular those with high
momenta) which are empty in mean field.
As a single-particle basis the relevant bound states of the Hartree-Fock
term are included and for states at positive energy plane waves are
employed with corresponding single-particle energies.
These plane waves are properly orthogonalized to the bound states (if present)
and enclosed in a box of sufficiently large radius to allow a convenient
discretization\cite{borro}.
The real part of the self-energy is obtained by using dispersion
relations relevant for these two self-energy contributions.
To avoid double counting for the real part, the corresponding
second order term calculated in nuclear matter at the original
starting energy is subtracted.

In the present work the OBEPC potential is used as a realistic
nucleon-nucleon interaction\cite{machl}.
Whereas the influence of short-range correlations is carefully
considered in this work, no attempt is made to treat the coupling
to the very low-lying two-particle-one-hole and two-hole-one-particle
states in an adequate way.
Attempts at such a treatment can be found in
Refs.~\cite{domit,brand,rijsd,skou1,skou2} (see also Ref.~\cite{dickh}).
To obtain the nucleon hole spectral function one needs to solve the Dyson
equation for the single-particle propagator
\begin{equation}
g_{lj}(k_1,k_2;E) = g^{(0)}_{lj}(k_1,k_2;E)
+ \sum_{k_3,k_4} g^{(0)}_{lj}(k_1,k_3;E) \Sigma_{lj}(k_3,k_4;E)
g_{lj}(k_4,k_2;E) ,
\label{eq:dyson}
\end{equation}
where $g^{(0)}$ refers to the Hartree-Fock propagator and $\Sigma_{lj}$
represents the real and imaginary part of the irreducible self-energy
calculated from the second order terms discussed above. For the
energies $E$ of interest here (below $\epsilon^-_F=E^A_0
-E^{A-1}_0$), the solutions of Eq.~(\ref{eq:dyson}) are insensitive to
the discretization of the momentum integrals, if the radius of the box,
which determines the grid of momenta $k_{i}$, is sufficiently large.
The spectral function for hole strength is obtained from the
diagonal matrix element of $g_{lj}$ by taking the imaginary part and
dividing by $\pi$.
In the present work the solution of the Dyson equation yields
discrete solutions corresponding to the $p{1\over 2}$ ground
state as well as the first $p{3\over 2}$ excited state
of the $A = 15$ system.
These discrete quasihole solutions are obtained by solving the
eigenvalue problem corresponding to Eq.~(\ref{eq:dyson}).
The eigenvector corresponding to these discrete states yields
the quasihole wave function in momentum space,
which still needs to be normalized by the spectroscopic factor
\begin{equation}
\mid w^{n}_{\alpha_{qh}} \mid^2 =
\bigg( {1-{\partial \Sigma(\alpha_{qh},\alpha_{qh};E) \over
\partial E} \bigg|_{\epsilon_{qh}}} \bigg)^{-1} ,
\label{eq:qhs}
\end{equation}
where $\alpha_{qh}$ corresponds to the quasihole single-particle
quantum numbers and the self-energy, which is real at the
quasihole energies $\epsilon_{qh}$, is calculated for these quantum
numbers\cite{dickh}.
The quasihole energies obtained in the present work yield -17.9 MeV
for the $p{3\over 2}$ and -14.1 MeV for the $p{1\over 2}$ state,
respectively.
The results for the strength of the quasihole poles is
0.89 for the $p{1\over 2}$ state and 0.91 for the
$p{3\over 2}$ state, respectively.
These numbers can be compared with experimentally determined
spectroscopic factors which have recently been determined at the
NIKHEF facility\cite{nikhe}.
Although the present theoretical result overestimates the experimental
result by about 0.2, it is clear that a considerable renormalization
of the strength is to be expected due the coupling of
the single-hole states to the low-lying collective excitations,
which are not treated in this work.
Instead, one should view the quasihole strength that is obtained here,
to be the result of the influence of short-range
correlations\cite{borro}.

The square of the quasihole wave function for the $p{1\over 2}$ state
(normalized to the spectroscopic factor, see Eq.~(\ref{eq:qhs}) is
shown in Fig.~\ref{fig:qpwf} as the full line.
For comparison the result for the Hartree-Fock wave function is shown
as the dashed line.
{}From the comparison one can infer that at the quasihole energies
no substantial change in the wave functions occur and that the Hartree-Fock
wave function is a good approximation.
It should further be noted that the wave function of a Woods-Saxon
potential, which is constructed as the local equivalent of the
Hartree-Fock potential\cite{borro}, is indistinguishable from the
Hartree-Fock wave function.
This suggests that the explicit inclusion of short-range correlations
does not lead to the strong suppression of the wave function
in the interior of the nucleus as has been implied by
Refs.~\cite{mawam,mahau}.
Again it should be emphasized that the coupling to the low-lying collective
excitations may lead to additional changes in the quasihole wave function.
It is unlikely, however, that these changes will involve the high-momentum
part of the wave function, which is studied in this work.
The calculation of the natural orbitals for the $p$-states does not yield
much new information.
With the emphasis on short-range correlations in the present work,
the diagonalization of the
one-body density matrix yields for its largest eigenvalue (0.93 for the
$p{3\over 2}$ orbital) a wave function
which is practically identical to the quasihole wave function
(which has strength 0.91).

The hole spectral function in the continuum is obtained from the
imaginary part of the single-particle propagator by expressing
the propagator in terms of the reducible self-energy
$\Sigma^R_{lj}$.
The reducible self-energy can be obtained from the irreducible
self-energy by a complex matrix
inversion, which solves an equation similar to Eq.~(\ref{eq:dyson}).
The results at three different energies are shown in Fig.~\ref{fig:npe12}.
The long-dashed curve corresponds to -50 MeV, the full curve
to -100 MeV, and finally the short-dashed curve to -250 MeV.
{}From these results it is clear that an important change in
the momentum content of the single-particle strength occurs with
increasing excitation energy in the $A = 15$ system.
At higher excitation energy one finds more high momentum
components.
Moreover, these high momentum components are not observed
in the quasihole states.
This can be concluded from Fig.~\ref{fig:np12} where the contribution
to the momentum distribution of the quasihole state (dashed line)
is compared with the total $p{1\over 2}$ momentum distribution
including the contribution of the continuum (full line).
This requires the integration of the continuum hole strength
for each $k$ according to Eq.~(\ref{eq:nljke}).
This result for $^{16}{\rm O}$ is very similar to the observation
for $^3{\rm He}$ made in Ref.~\cite{tadok}, where the contribution
of the ground state to ground state transition exhibits also
very few high-momentum components.

To understand this result, it is important to recall
that the appearance of high-momentum components at a certain
energy in the $A-1$ system is related to the self-energy contribution
containing two-hole-one-particle states at this energy.
{}From energy conservation it is then clear that at low energy
it is much harder to find such states with a high-momentum particle
than at high energy.
This same feature is observed in nuclear matter where the peak
of the single-particle spectral function for momenta above $k_F$
increases in energy as $k^2$\cite{ciofb}.
As a result, the hole strength in nuclear matter as a function
of momentum shows the same tendency as the result shown in
Fig.~\ref{fig:npe12}\cite{gearh}, {\it i.e.} higher momenta become more
dominant
at higher excitation energy.

The additional $p{1\over 2}$-strength in the continuum,
which is illustrated in Fig.~\ref{fig:npe12}, integrated over
$k$ yields 4\% of single-particle strength.
Together with the quasihole strength this leads to an occupation of
$p{1\over 2}$ quantum numbers which is less than one.
This implies that higher
waves like $d$ and $f$ will be partially occupied.
Results for these higher waves (and the $s$-wave) will be discussed elsewhere
together with results for the average separation energy,
Koltun sum rule etc.\cite{mueth}.

In conclusion, it has been shown in this work that the influence of
high-momentum components in the quasihole wave function is of minor
importance.
By calculating the complete energy dependence of the $p$-wave hole
spectral function it has been demonstrated that the presence of high-momentum
components in the nuclear ground state will only show up unambiguously
at high excitation energy when the $(e,e'p)$ reaction is employed.

This work has been supported by the German DFG (Mu 705/3)
and the US NSF under Grant
PHY-9002863.

\begin{figure}
\caption{Square of the quasihole wave function for the $p {1\over 2}$ state
in $^{16}{\rm O}$
(full curve), normalized to the spectroscopic factor according to Eq.~(5),
compared to the Hartree-Fock result (dashed curve).}
\label{fig:qpwf}
\end{figure}
\begin{figure}
\caption{The $p{1\over 2}$ spectral function as a function of momentum at
fixed
energies corresponding to -50, -100, and -250 MeV.
The results demonstrate the increasing importance of high-momentum components
with higher excitation energy in $A-1$ system (more negative energy).}
\label{fig:npe12}
\end{figure}
\begin{figure}
\caption{Total $p{1\over 2}$ momentum distribution (full curve)
compared to quasihole contribution (dashed curve).}
\label{fig:np12}
\end{figure}


\begin{references}
\bibitem{zabol} J.G. Zabolitzky and W. Ey, Phys.\ Lett.\ {\bf 76 B},
527 (1978).
\bibitem{orden} J.W. Van Orden, W. Truex, and M.K. Banerjee, Phys.\
Rev.\ C {\bf 21}, 2628 (1980).
\bibitem{benha} O. Benhar, C. Ciofi degli Atti, S. Liuti, and
G. Salm\`e, Phys.\ Lett.\ {\bf 177 B}, 135 (1986).
\bibitem{ji} X. Ji and J. Engel, Phys.\ Rev.\ C {\bf 40}, R497 (1989).
\bibitem{string} S. Stringari, M. Traini, and O. Bohigas, Nucl.\ Phys.\
{\bf A516}, 33 (1990).
\bibitem{tadok} S. Tadokoro, T. Katyama, Y. Akaishi, and H. Tanaka,
Prog.\ Theor.\ Phys.\ {\bf 78}, 732 (1987).
\bibitem{ciofi} C. Ciofi degli Atti, E. Pace, and G. Salm\`e, Phys.\
Lett.\ {\bf 141 B}, 14 (1984).
\bibitem{lapik} NIKHEF-K Proposal NR: 91-E19, L. Lapik\'as spokesperson.
\bibitem{lewar} D.S. Lewart, V.R. Pandharipande, and S.C. Pieper,
Phys.\ Rev.\ B {\bf 37}, 4950 (1988).
\bibitem{mawam} Z.Y. Ma and J. Wambach, Phys.\ Lett.\ {\bf 256 B}, 1 (1991).
\bibitem{mahau} C. Mahaux and R. Sartor, Adv.\ Nucl.\ Phys.\
{\bf 20}, 1 (1991).
\bibitem{ramos} A. Ramos, A. Polls, and W.H. Dickhoff, Nucl\ Phys.\
{\bf A503}, 1 (1989).
\bibitem{benff} O. Benhar, A. Fabrocini, and S. Fantoni, Nucl.\ Phys.\
{\bf A505}, 267 (1989).
\bibitem{dickh} W.H. Dickhoff and H. M\"uther, Rep.\ Prog.\
Phys.\ {\bf 55}, 1947 (1992)
\bibitem{diepe} A.E.L. Dieperink, T. De Forest Jr., I. Sick, and
R.A. Brandenburg, Phys.\ Lett.\ {\bf 63 B}, 261 (1976).
\bibitem{borro} M. Borromeo, D. Bonatsos, H. M\"uther, and A. Polls,
Nucl.\ Phys.\ {\bf A539}, 189 (1992).
\bibitem{machl} R. Machleidt, Adv.\ Nucl.\ Phys.\ {\bf 19}, 1 (1989).
\bibitem{domit} W.H. Dickhoff, P.P. Domitrovich, A. Polls, and A. Ramos,
in {\it Condensed Matter Theories}, Vol.5, V.C. Aguilera-Navarro ed.
(Plenum, New York, 1990) p 275;
P.P. Domitrovitch, thesis Washington University (1991) and in preparation.
\bibitem{brand} M.G.E. Brand, G.A. Rijsdijk, F.A. Muller, K. Allaart,
and W.H. Dickhoff, Nucl.\ Phys.\ {\bf A531}, 253 (1991).
\bibitem{rijsd} G.A. Rijsdijk, K. Allaart, and W.H. Dickhoff, Nucl.\
Phys.\ {\bf A550}, 159 (1992).
\bibitem{skou1} H. M\"uther and L.D. Skouras, Phys.\ Lett.\ {\bf 306 B},
306 (1993).
\bibitem{skou2} H. M\"uther and L.D. Skouras, Nucl.\ Phys.\ {\bf A},
in press.
\bibitem{nikhe} NIKHEF Annual Report 1991.
\bibitem{ciofb} C. Ciofi degli Atti, S. Simula, L.L. Frankfurt, and
M.I. Strikman, Phys.\ Rev.\ C{\bf 44}, R7 (1991).
\bibitem{gearh} C.C. Gearhart, W.H. Dickhoff, A. Polls, and A. Ramos,
to be published.
\bibitem{mueth} H. M\"uther, A. Polls, and W.H. Dickhoff,
to be published.
\end{references}
\end{document}